\documentclass[conference]{IEEEtran}

\usepackage{booktabs}
\usepackage{amsmath}
\usepackage{amsthm}
\usepackage{multirow}
\usepackage{color,cite}
\usepackage{subfigure}
\usepackage{mathrsfs}
\usepackage{amssymb}
\usepackage{bm}
\usepackage{graphicx}
\usepackage{amsfonts}
\usepackage{algorithm}
\usepackage{algorithmic}
\usepackage{epstopdf}

\begin{document}

\bibliographystyle{IEEEtran} 

\title{Deep Learning Assisted mmWave Beam Prediction\\ with Prior Low-frequency Information
 \vspace{-3mm}}

\author{
\IEEEauthorblockN{Ke Ma$^{\rm 1}$, Dongxuan He$^{\rm 1}$, Hancun Sun$^{\rm 1}$, Zhaocheng Wang$^{\rm 1, \rm 2}$} %

\IEEEauthorblockA{$^{\rm 1}$Tsinghua National Laboratory for Information Science
 and Technology, \\
 Department of Electronic Engineering, Tsinghua University, Beijing 100084, China\\
 $^{\rm 2}$Tsinghua Shenzhen International Graduate School, Shenzhen 518055, China\\
 (\textbf{Accepted by IEEE ICC 2021})}
 \vspace{-8.5mm}
}

\maketitle

\begin{abstract}
Huge overhead of beam training poses a significant challenge to mmWave communications.
To address this issue, beam tracking has been widely investigated whereas existing methods are hard to handle serious multipath interference and non-stationary scenarios.
Inspired by the spatial similarity between low-frequency and mmWave channels in non-standalone architectures, this paper proposes to utilize prior low-frequency information to predict the optimal mmWave beam, where deep learning is adopted to enhance the prediction accuracy. Specifically, periodically estimated low-frequency channel state information (CSI) is applied to track the movement of user equipment, and timing offset indicator is proposed to indicate the instant of mmWave beam training relative to low-frequency CSI estimation. Meanwhile, long-short term memory networks based dedicated models are designed to implement the prediction. Simulation results show that our proposed scheme can achieve higher beamforming gain than the conventional methods while requiring little overhead of mmWave beam training.
\end{abstract}

\IEEEpeerreviewmaketitle

\vspace{-.2mm}
\section{Introduction}\label{sec:intro}
\vspace{-.0mm}

 Millimeter-wave (mmWave) communication has been one of the crucial technologies in fifth-generation (5G) systems due to its sufficient bandwidth resources. To compensate the severe pathloss of the mmWave signals, large antenna arrays are equipped at both base stations (BSs) and user equippments (UEs) to establish directional links, where beam training is performed to search the optimal direction. However, the optimal beam switches rapidly in mobile scenarios, which requires frequent beam training and brings huge training overhead.



 Beam tracking has been extensively investigated to reduce the training overhead, where the conventional solutions focus on tracking the change of angle-of-departures (AoDs) and angle-of-arrivals (AoAs).
 \cite{Ref:EKF1} and \cite{Ref:PF} proposed to use one beam pair to track the dominant path by the Extended Kalman Filter (EKF) and the Particle Filter (PF), respectively.
 However, low signal-to-noise ratio (SNR) may lead to significant misconvergence since slow variations of AoD/AoA in mobile scenarios are not considered. 
 To address the problem, \cite{Ref:EKF2} introduced angular velocity into the state estimation of EKF and utilized three received reference signals to jointly mitigate the influence of noise. However, the transmission of reference signals induces extra overhead.
 Moreover, the angles of paths in the local cluster formed by the reflection of scatterers adjacent to UE (e.g., bodies and clothes) are closely near the dominant path \cite{Ref:LOCAL_CLUSTER}, causing serious multipath interference, which is inconsistent with the assumption shared by \cite{Ref:EKF1,Ref:PF,Ref:EKF2} that only one single path falls into the main beam direction.

 Despite the widespread applications in beam tracking, the state-of-the-art methods are difficult to handle non-stationary scenarios, e.g., vehicles with high and time-varying speeds. Consequently, deep learning has been introduced to extract highly nonlinear correlations and reduce the overhead of beam training. In \cite{DL1}, long-short term memory networks (LSTMs) were adopted to infer optimal mmWave beams at the target BS based on the channel state information (CSI) of the source mmWave BS. However, the estimation of mmWave CSI may lead to huge overhead due to large antenna numbers. Furthermore, \cite{DL2} utilized deep reinforcement learning to track the dominant path based on the selected beam pair.
 Nevertheless, the measurement of single beam pair is sensitive to noise and multipath interference.

Usually, mmWave antennas are deployed on existing low-frequency BSs in non-standalone (NSA) architectures. Field experiments in \cite{LOWFRE1,LOWFRE2} demonstrated that low-frequency and mmWave channels share similar spatial features including AoDs and AoAs in NSA architectures, which provides the feasibility for reducing the training overhead by using low-frequency information. \cite{LOWFRE1} proposed to search mmWave beams within the 3dB bandwidth of the optimal sub-6GHz beam, whereas exhaustive search in the candidate range still brings huge overhead.
 In \cite{LOWFRE3} and \cite{LOWFRE4}, the optimal mmWave beam was predicted based on the instantaneous low-frequency CSI.
 However, low-frequency CSI estimation and mmWave beam training are not guaranteed to be synchronized, which restricts its feasibility.

 In order to further reduce the overhead of beam training, we propose to utilize prior low-frequency information to predict the optimal mmWave beam, where deep learning is adopted to enhance prediction accuracy in severe multipath interference and non-stationary scenarios. Specifically, the NSA architecture including low-frequency and mmWave links is investigated, where the spatial features of low-frequency and mmWave channels are assumed to be similar. Periodically estimated low-frequency CSI is used to track the movement of UE and predict the optimal mmWave beam direction without additional overhead.
 To handle the timing asynchronization between low-frequency and mmWave links, we propose timing offset indicator to indicate the instant of mmWave beam training relative to low-frequency CSI estimation. Furthermore, LSTM based dedicated models are designed to implement the prediction. Simulation results demonstrate that our proposed scheme can achieve higher beamforming gain compared with its conventional counterparts, which ensures high mmWave communication quality with little overhead of beam training.



\section{System Model}\label{sec:sys}

 The NSA architecture consisting of the low-frequency link and the mmWave link is considered, where multiple input multiple output (MIMO) is adopted by both links.
 For simplicity, here we assume the two-dimensional (2D) channel model such that only azimuth angles are considered.
 Since the line-of-sight (LOS) path enjoys low attenuation, we consider the LOS scenario.

 For the low-frequency link, a time-varying geometric downlink channel model is established, which consists of a LOS path and $C$ clusters including one local cluster. Assuming that BS and UE are equipped with $M$ and $N$ antennas, the low-frequency channel $\mathbf{H}_{k} \in \mathbb{C}^{N \times M}$ at time slot $k$ is written as
\begin{equation}
\begin{split}
\mathbf{H}_k = & \sqrt{\frac{MN}{\rho_{\text{LOS}}}} \alpha_{\text{LOS}} \mathbf{a}_{\text{RX}} ( \theta_{\text{LOS}} ) \mathbf{a}_{\text{TX}}^{\text{H}} ( \phi_{\text{LOS}} ) + \\ & \sum_{c=1} ^{C}\sqrt{\frac{MN}{\rho_{c}}} \sum_{l=1} ^{L_{c}} \frac{\alpha_{c,l}}{\sqrt{L_{c}}} \mathbf{a}_{\text{RX}} (\theta_{c} + \theta_{c,l} ) \mathbf{a}_{\text{TX}}^{\text{H}} ( \phi_{c} + \phi_{c,l} ),
\end{split}
\end{equation}
where the time slot subscript $k$ is omitted. The $c$-th cluster containing $L_{c}$ paths has pathloss $\rho_{c}$, AoA $\theta_{c}$ and AoD $\phi_{c}$, while $\alpha_{c,l}$, $\theta_{c,l}$, $\phi_{c,l}$ denote the complex gain, AoA offset, AoD offset corresponding to the $l$-th path in the $c$-th cluster. Besides, the variables with subscript $\text{LOS}$ define corresponding parameters for the LOS path.
$\mathbf{a}_{\text{TX}}(\cdot) \in \mathbb{C}^{M \times 1}$ and $\mathbf{a}_{\text{RX}}(\cdot) \in \mathbb{C}^{N \times 1}$ denote the antenna response vectors of BS and UE. We assume that uniform linear arrays (ULAs) are adopted at both BS and UE sides, and the antenna response vector at BS side is expressed as
\begin{equation}
\mathbf{a}_{\text{TX}} ( \phi ) = \sqrt{\frac{1}{M}} [1,e^{j\frac{2\pi d_{\text{TX}} \sin\phi}{\lambda}},..., e^{j\frac{2\pi (M-1) d_{\text{TX}} \sin\phi}{\lambda}}]^{\text T},
\end{equation}
where $d_{\text{TX}}$ is the antenna spacing and $\lambda$ is the wavelength. The antenna response vector of UE can be obtained in a similar manner.

In order to obtain significant AoD/AoA information, we transform the channel to the angular domain as $\mathbf{H}_{k}^\text{ag} \in {\mathbb{C}^{N \times M}}$ by performing discrete Fourier transform (DFT) \cite{AD}, i.e.,
\begin{equation}
\mathbf{H}_{k}^\text{ag} = \mathbf{F}_{N} \mathbf{H}_k \mathbf{F}_{M}^{\text{H}},
\end{equation}
where $\mathbf{F}_{N} \in {\mathbb{C}^{N \times N}}$ and $\mathbf{F}_{M} \in {\mathbb{C}^{M \times M}}$ are DFT matrices.

Similarly, the mmWave channel is based on the time-varying geometric channel model. To distinguish from the low-frequency counterpart, the variables corresponding to the mmWave band are overlined. We assume phase shifter based analog beamforming is appied in the mmWave link, where $\mathbf{\overline{f}} \in \mathbb{C}^{\overline{M} \times 1}$ is denoted as the transmitting beam of BS and $\mathbf{\overline{w}} \in \mathbb{C}^{\overline{N} \times 1}$ is denoted as the receiving beam of UE. The transmitting beam and the receiving beam are selected from the predefined codebooks $\mathcal{\overline{F}}$ and $\mathcal{\overline{W}}$, respectively. In order to obtain the optimal beam pair $\{\mathbf{\overline{f}}^{*}_k , \mathbf{\overline{w}}^{*}_k\}$, conventional beam training sweeps all candidate beam pairs to select the beam pair with the maximum beamforming gain, which can be formulated as
\begin{equation}\label{beam_pair_training}
\{\mathbf{\overline{f}}^{*}_k , \mathbf{\overline{w}}^{*}_k\}=\mathop{\arg\max}\limits_{\mathbf{\overline{f}} \in \mathcal{\overline{F}},\mathbf{\overline{w}} \in \mathcal{\overline{W}}} | \mathbf{\overline{w}}^{\rm H} \mathbf{\overline{H}}_{k} \mathbf{\overline{f}}|.
\end{equation}

We assume that the spatial features of low-frequency and mmWave channels are similar in the NSA architecture. In particular, since both channels share the same propagation environment and surrounding scatterers, low-frequency and mmWave channels are considered to have identical LOS AoD/AoA and analogous cluster AoDs/AoAs \cite{LOWFRE3, LOWFRE5}. For AoDs, the correlation can be formulated as
\begin{equation}
\begin{split}
\overline{\phi}_{\text{LOS}} = &~ \phi_{\text{LOS}}, \\
\overline{\phi}_{c} = &~ \phi_{c} + n_\phi, 1 \leq c \leq C,
\end{split}
\end{equation}
where $n_\phi$ denotes additional white Gaussian noise (AWGN) with variance $\sigma_\phi^2$, and the correlation of AoAs can be expressed in the similar manner.
However, due to more serious pathloss of the mmWave band, mmWave channels have smaller AoD/AoA spread and delay spread, larger Ricean K factor (i.e., power ratio of the LOS path to other paths) and shadow fading compared with low-frequency channels \cite{LOWFRE5, RK}.

\section{Deep Learning Assisted \\ mmWave Beam Prediction}\label{S3}

\begin{figure*}
\centering
\includegraphics[width=0.9\textwidth]{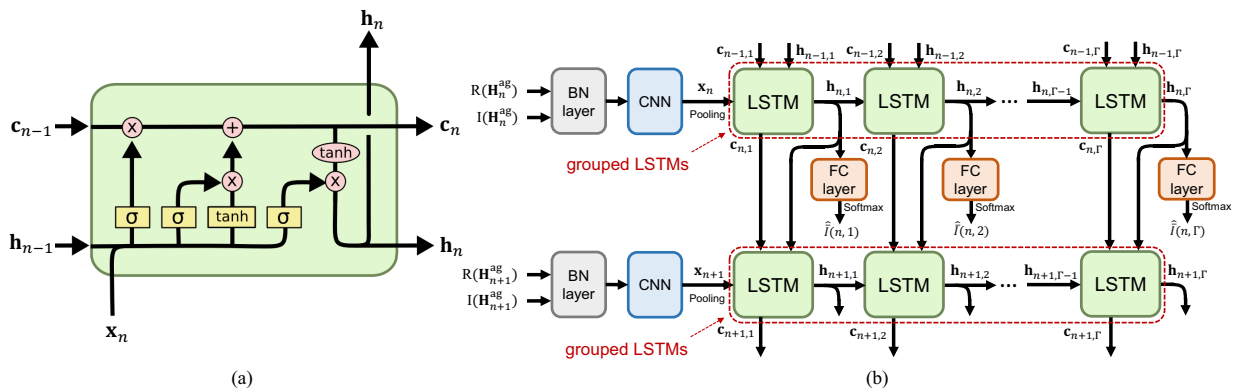}
\caption{(a) Basic structure of LSTM. (b) Proposed deep learning model design, where $\widehat{\overline{I}}(n, \gamma)$ is the predicted mmWave beam for the $\gamma$-th instant after the $n$-th low-frequency CSI estimation. $\Gamma$ LSTMs together with following FC layers share the same structures and parameters, which are grouped and have one-to-one correspondence with prediction results in time order. }
\vspace{-3mm}
\label{model}
\end{figure*}

\subsection{Problem Formulation}

In this paper, we propose to predict the optimal mmWave beam by using prior low-frequency information.
For simplicity, here we investigate the prediction of the optimal mmWave transmitting beam at BS side, whereas the same method can be easily extended to predict the optimal receiving beam.
Since low-frequency CSI is periodically estimated for the operation of the low-frequency link itself \cite{LOWFRE1}, low-frequency CSI can be used as the prediction input without additional overhead. Considering that low-frequency CSI estimation and mmWave beam training are not guaranteed to be synchronized, we propose to adopt the sequence of periodically estimated low-frequency CSI to track the movement of UE and predict the optimal mmWave beam when mmWave beam training is required. To indicate the instant of mmWave beam training relative to low-frequency CSI estimation last time, timing offset indicator $\overline{\eta}$ is proposed and can be calculated as
\begin{equation}
\overline{\eta} = \frac{\overline{t}-t_n}{T},
\end{equation}
where $T$ denotes the period of low-frequency CSI estimation and $t_n$ denotes the instant of the $n$-th low-frequency CSI estimation. $\overline{t}$ is the instant of mmWave beam training, which satisfies $t_n \leq \overline{t} < t_n + T$.

Since the number of candidate beams at BS side is limited, the beam prediction can be formulated as a multi-classification problem, where each category corresponds to a specific mmWave beam in the codebook $\mathcal{\overline{F}}$. Therefore, the prediction model can be represented by the classification function $f(\cdot)$ as
\begin{equation}
\overline{I}_n(\overline{t})=f(\mathbf{H}^{\text{ag}}_{n-m+1},\mathbf{H}^{\text{ag}}_{n-m+2},...,\mathbf{H}^{\text{ag}}_{n}, \overline{\eta}),
\end{equation}
where $m$ prior angular-domain low-frequency CSI is applied to predict the optimal mmWave transmitting beam index $\overline{I}_n(\overline{t}) \in \{ {1,2,...,\overline{N}_{\text{BS}}} \}$, and $\overline{N}_{\text{BS}}$ denotes the codebook size, i.e., $\overline{N}_{\text{BS}} = |\overline{\mathcal{F}}|$. Because conventional methods are difficult to handle non-stationary scenarios and serious multipath interference, deep learning models which enjoy strong abilities to learn complex nonlinear correlations are introduced to implement the prediction. Besides, the prediction is proposed to be mainly performed at BS side, whose strong computational capability can ensure low prediction delay.

The proposed scheme consists of two stages, i.e., the training stage and the predicting stage.
In the training stage, training samples are collected to train the deep learning model.
For each UE, periodically estimated low-frequency CSI and optimal mmWave transmitting beam indices obtained by conventional beam sweeping are packed in time order separately and labeled by corresponding instants, which forms a training sample.
The optimal mmWave transmitting beam index is used as the classification label, where BS calculates its timing offset indicator and utilizes corresponding prior low-frequency CSI as prediction input to train the model.
After the deep learning model is well-trained by sufficient training data, the proposed scheme comes to the predicting stage.
When mmWave beam training is required, BS leverages the well-trained deep learning model to predict the optimal transmitting beam for the mmWave link according to prior low-frequency CSI and corresponding timing offset indicator, such that huge beam training overhead can be mitigated efficiently.

The deep learning model trained by field data can adapt well to surrounding propagation environments. Once the environment varies, BS can collect new training data and update the deep learning model in the predicting stage to follow the changes in time.


\subsection{Deep Learning Model Design}

Specifically, we use LSTM as the prediction model due to its outstanding performance in sequence feature extractions.
The basic structure of LSTM is shown in Fig. 1(a), where the input of current time slot $\mathbf{x}_n$ together with the cell state $\mathbf{c}_{n-1}$ and the output $\mathbf{h}_{n-1}$ of former time slot are fed to the LSTM at $n$-th time slot, so that LSTM could learn the features from prior inputs.

However, directly leveraging conventional LSTM is not suitable in this case, because the instant of mmWave beam training is not synchronized to low-frequency CSI estimation but LSTM implicitly assumes equally sampled outputs. To tackle the problem, we propose to use LSTM to ``\emph{interpolate}" the optimal mmWave beams at several instants between two times of low-frequency CSI estimation, where the predicted beam with the shortest time interval from mmWave beam training is used as the corresponding optimal beam. The structure of proposed deep learning model design is shown in Fig. 1(b).

In order to better learn sequence features, input low-frequency CSI is preprocessed before being fed to LSTM. Firstly, since the CSI matrix is complex-valued which cannot be used as LSTM input directly, the input is divided to two real-valued feature channels, including the real part $\mathrm{R}(\mathbf{H}^{\text{ag}}_n)$ and the imaginary part $\mathrm{I}(\mathbf{H}^{\text{ag}}_n)$. Secondly, due to huge dynamic range of CSI values, batch normalization (BN) is performed to transform the input to the standard distribution with expectation 0 and variance 1 \cite{BATCH_NORM}. Finally, convolutional neural networks (CNNs) and pooling layers are adopted to extract preliminary features from the input CSI matrix.

Once preprocessed, the extracted features of CNN are fed to LSTM. We define the interpolation factor of beam prediction as $\Gamma$, which means the optimal mmWave beams corresponding to $\Gamma$ instants between two times of low-frequency CSI estimation are predicted. Uniform sampling is applied for simplicity, where $\Gamma$ instants contain $t_n + \frac{2\gamma - 1}{2\Gamma}T,~ 1 \leq \gamma \leq \Gamma$. Therefore, we adopt the predicted result closest in time to mmWave beam training as the approximation of the corresponding optimal mmWave beam, which can be expressed as
\begin{equation}
\begin{split}
\widehat{\overline{I}}_n(\overline{t}) &= g(\mathbf{H}^{\text{ag}}_{n-m+1},\mathbf{H}^{\text{ag}}_{n-m+2},...,\mathbf{H}^{\text{ag}}_{n}, \gamma(\overline{t})), \\
\gamma(\overline{t}) &= \mathop{\arg\min}\limits_{\gamma} |\overline{\eta} - \frac{2\gamma - 1}{2\Gamma} |,~ \gamma \in \{ 1, 2, ..., \Gamma \},
\end{split}
\end{equation}
where $\widehat{\overline{I}}_n(\overline{t})$ is the predicted optimal mmWave beam at instant $\overline{t}$, and $g(\cdot)$ denotes the proposed deep learning model.
Because the predicted results share the same time interval, grouped LSTMs is proposed to predict $\Gamma$ beams after each low-frequency CSI estimation, where the sequence of $\Gamma$ LSTMs with the same structures and parameters is included and each LSTM corresponds to the predicted beam in time order, so that model storage overhead is efficiently reduced. The first LSTM utilizes current low-frequency CSI as model input, while the $\gamma$-th LSTM utilizes the output of $(\gamma - 1)$-th LSTM as model input.

To predict the optimal beam from all candidate beams, the fully-connected (FC) layer is added after LSTM to implement the size transformation from the output of LSTM to the number of candidate beams, followed by a softmax activation layer which normalizes the output into probabilities, which can be written as
\begin{equation}
\widehat{W}_{i}(\overline{t})=\text{softmax}(\mathbb{W}_{i}\mathbf{h}(n, \gamma(\overline{t}))+b_{i}),i\in\{1,2,...,\overline{N}_{\text{BS}}\},
\end{equation}
where $\mathbf{h}(n, \gamma(\overline{t}))$ is the output vector of LSTM, $\mathbb{W}_{i}$ and $b_{i}$ denote the weights and bias of the FC layer corresponding to the $i$-th output. Similarly, all FC layers after $\Gamma$ LSTMs share the same structures and parameters. Finally, the beam with the maximum proability is selected, i.e.,
\begin{equation}
\widehat{\overline{I}}_n(\overline{t}) = \mathop{\arg\max}\limits_{i\in\{1,2,...,\overline{N}_{\text{BS}}\}}\widehat{W}_{i}(\overline{t}).
\end{equation}

Note that the predicted probabilities provide a judgement of beam qualities, i.e., the beam with larger probability is predicted to enjoy higher quality compared with the counterpart with smaller probability. If the predicted optimal beam fails, we can try other beams based on predicted probabilities, which could further reduce the overhead of mmWave beam training.

Cross entropy loss is one of the evaluation metrics widely used in classification problems, which is utilized to train our proposed model. Mathematically, it can be expressed as
\begin{equation}
\text{loss}(\overline{t}) = -\sum\limits_{i=1} \limits^{\overline{N}_{\text{BS}}} W_{i} (\overline{t}) \log(\widehat{W}_{i} (\overline{t}) ),
\end{equation}
where $W_i(\overline{t})$ is the actual optimal mmWave beam at instant $\overline{t}$ expressed by one-hot encoding. For one training sample, the overall loss is the average of prediction losses at all instants that the actual optimal mmWave beam is obtained as the classification label.

\section{Simulation Analysis}

\subsection{Simulation Setup}

We consider an NSA wireless system consisting of a sub-6GHz link and a mmWave link, where cell radius is $100$m, and UE movement contains two scenarios, i.e., the stationary scenario and the non-stationary scenario. In the stationary scenario, UE moves with speed $v_\text{s}\in[8, 12]$m/s and no acceleration. In the non-stationary scenario, UE has speed $v_{\text{ns}}\in[25, 30]$m/s and acceleration $a_{\text{ns}}\in[-5, 5]\text{m/s}^2$. Besides, the direction of all movements are randomly generated in $[0, 2\pi]$. The period of low-frequency CSI estimation is $200$ms, while the instant of mmWave beam training is randomly generated. In addition, the number of mmWave candidate transmitting beams $\overline{N}_{\text{BS}}$ is $32$.

In our simulation, we extend the COST2100 channel model to time-varying dual-band scenarios, where one dominant LOS path together with far clusters and local clusters are considered \cite{COST2100}. As we mentioned in Sec. II, the LOS paths of low-frequency and mmWave bands share the same AoD/AoA. Far clusters are reflected by surrounding scatterers (e.g., buildings), where the perturbation given by white Gaussian noise with variance $\sigma_x^2 = \sigma_y^2 = (0.8\text{m})^2$ is added to low-frequency scatterer location $[x, y]$ for generating its mmWave counterpart $[\overline{x}, \overline{y}]$, so that cluster AoD/AoA offset $n_\phi,n_\theta$ is simulated. Differently, local clusters are reflected by scatterers adjacent to UE, thus it is assumed that these scatterers are randomly distributed around UE location. Besides, the COST2100 channel model defines visible region corresponding to each cluster, i.e., the area where the cluster exists. Since the mmWave band suffers more serious pathloss, the size of mmWave visible regions is smaller than low-frequency visible regions.

\begin{table}
\begin{center}
\caption{Dual-band channel parameters.}
\setlength{\tabcolsep}{2mm}
\begin{tabular}{ccccccc}
\toprule[0.8pt]
Channel & Low-frequency & mmWave  \\
\toprule[0.8pt]
Center frequency $f_c / \overline{f}_c$                        &$3.5$GHz              &$28$GHz                 \\
Far cluster number $c_f / \overline{c}_f$                      &$15$                  &$15$                 \\
Path number within one cluster $L_c / \overline{L}_c$        &$20$                  &$20$                      \\
Visible region radius $r_v / \overline{r}_v$        &$50$m                  &$30$m                      \\
Cluster AoD spread $\Delta \phi / \Delta \overline{\phi}$        &$6^\circ$             &$1.9^\circ$           \\
Cluster delay spread  $\tau / \overline{\tau}$    &$10$ns             &$3$ns               \\
Shadow fading standard deviation $\sigma_{\text{SF}}/\overline{\sigma}_{\text{SF}}$   &$2$dB        &$4$dB                    \\
Antenna number $M \times N / \overline{M} \times \overline{N}$   &$8 \times 1$           &$32 \times 1$                     \\
\toprule[0.8pt]
\end{tabular}
\label{Tab:paramters}
\end{center}
\end{table}

\begin{table}
\begin{center}
\caption{Structure and parameters of proposed deep learning model.}
\setlength{\tabcolsep}{2mm}
\begin{tabular}{ccccccc}
\toprule[0.8pt]
Layer & Structure & Parameter  \\
\toprule[0.8pt]
Layer 1 &BN              &$f_{i}=2,f_{o}=2$                 \\
Layer 2 &CNN                              &$f_{i}=2,f_{o}=64$, stride $=3$, kernel size $ = 3$                 \\
Layer 3 &Activation                              &$f_{i}=64,f_{o}=64$, ReLU                  \\
Layer 4 &CNN                             &$f_{i}=64,f_{o}=256$, stride $=3$, kernel size $ = 1$                       \\
Layer 5 &Activation                              &$f_{i}=256,f_{o}=256$, ReLU                 \\
Layer 6 &Pooling             &$f_{i}=256,f_{o}=256$, max-pooling           \\
Layer 7 &LSTM             &$f_{i}=256,f_{o}=256$               \\
Layer 8 &FC        &$f_{i}=256,f_{o}=32$                    \\
Layer 9 &Activation   &$f_{i}=32,f_{o}=32$, softmax                     \\
\toprule[0.8pt]
\end{tabular}
\label{Tab:paramters}
\end{center}
\end{table}

Specific dual-band parameters are shown in Table I. For beam prediction, noise and multipaths may lead to wrong predicted results. We assume SNRs of both links are $20$dB because of low attenuation of the LOS path and focus on investigating the impact of multipaths here, where low-frequency Ricean K factor $R$ is selected from $\mathcal{R}=\{0,4,8,12,16,20\}$dB, while mmWave Ricean K factor is fixed $\overline{R}=20$dB.



The detailed structure and parameters of proposed deep learning model are shown in Table II, where $f_i$ and $f_o$ denote the numbers of input feature channels and output feature channels. We construct a dataset with 10,240 samples, where 80\% and 20\% of the dataset is used as the training set and the validation set, respectively. The model is trained for $40$ epochs in the training stage, where Adam optimizer based on back propagation algorithm is used to optimize model parameters \cite{ADAM}.

The EKF based beam tracking method in [1] and the deep learning based method without using prior information in [9] are used as our baselines. For proper comparison, we introduce angular velocity and angular acceleration into the state of EKF to handle mobile scenarios, and beam tracking is performed per $20$ms by utilizing low-frequency beams, where low-frequency candidate beam number equals antenna number, i.e., $N_{\text{BS}}=M$.

Two metrics including cross entropy loss and beamforming gain ratio are used for evaluating performance. Cross entropy loss is leveraged for tracking the training process of our proprosed scheme, while beamforming gain ratio $p$ is adopted to evaluate the quality of predicted beams, which is calculated as
\begin{equation}
p = \frac{1}{|\mathcal{V}|}\sum\limits_{v = 1}^{|\mathcal{V}|} \frac{|\mathbf{\overline{H}}_v \widehat{\overline{\mathbf{f}}}^*_v|}{|\mathbf{\overline{H}}_v \overline{\mathbf{f}}^*_v|},
\end{equation}
where $\mathcal{V}$ denotes the validation set, and $\widehat{\overline{\mathbf{f}}}^*_v$ is the predicted optimal transmitting beam. Receiving beams are omitted due to single antenna.








\subsection{Simulation Results}

We investigate the impact of interpolation factors $\Gamma$ on the loss of the validation set under Ricean K factor $R=8$dB, where the average results of three times of training are depicted in Fig. 2. It is obvious that the converged loss becomes smaller as $\Gamma$ increases from $1$ to $4$, since larger $\Gamma$ provides smaller approximation error. However, the convergence becomes slower and the converged loss is larger when $\Gamma=6$ compared with $\Gamma=4$ in both stationary and non-stationary scenarios, which is because sophisticated networks are more likely to overfit. Therefore, we apply $\Gamma = 4$ in the following simulations, which achieves good tradeoff between approximation errors and network complexities.

\begin{figure}
\centering
\includegraphics[width=.475\textwidth]{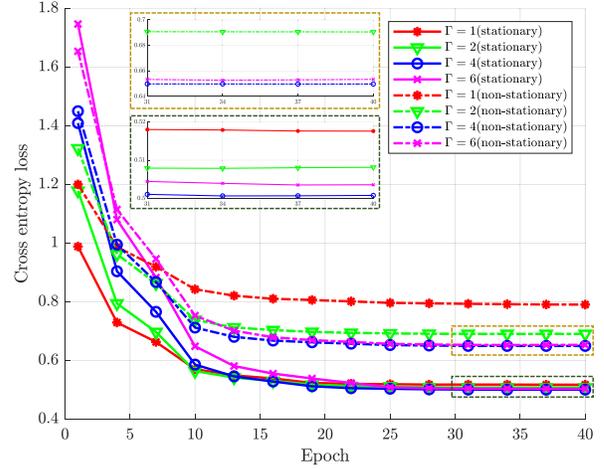}
\caption{Cross entropy loss comparison under different interpolation factors.}
\end{figure}

Next, the impact of prior low-frequency CSI length $m$ on beamforming gain ratio with fixed Ricean K factor $R=8$dB is investigated in Fig. 3. The result of baselines is stable since they do not rely on prior low-frequency CSI. In stationary scenarios, we can see that the performance of our proposed scheme surpasses EKF based method in [1] when $m=3$, which shows the proposed scheme can track UE movement more accurately than EKF. Furthermore, the performance of EKF based method degrades significantly under the non-stationary scenario, but our proposed scheme can extract robust nonlinear features from prior low-frequency CSI and achieves considerable performance enhancement. Besides, we can see that the performance of our proposed scheme converges when $m=5$ in both scenarios, which indicates that UE only needs to wait for at most $1$s to obtain stable mmWave beam alignment.

\begin{figure}
\centering
\includegraphics[width=.49\textwidth]{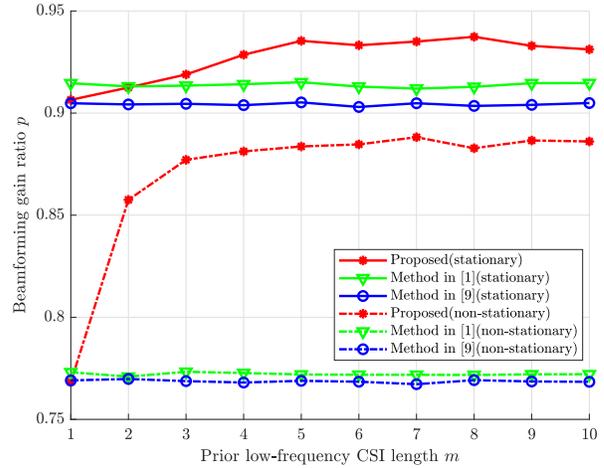}
\caption{Beamforming gain ratio comparison under different prior low-frequency CSI lengths.}
\end{figure}

Figure 4 shows the beamforming gain ratio at different instants from last low-frequency CSI estimation, where the instants are expressed by timing offset indicator $\overline{\eta}$ and Ricean K factor is fixed $R=8$dB. It can be seen that the proposed scheme enjoys better tracking ability than EKF based method in [1] in stationary scenarios. However, it is observed that the performance of baselines decreases rapidly with increasing $\overline{\eta}$ in non-stationary scenarios, since deep learning based method in [9] cannot track UE movement and EKF based method in [1] lacks the capability to handle highly nonlinear correlations. However, our proposed scheme still maintains at least 85\% beamforming gain ratio, which guarantees high channel quality at any instant. Besides, it is interesting to see that there is an obvious performance degradation of our proposed scheme when $\overline{\eta}=0.5$ in non-stationary scenarios, because its corresponding instant, i.e., $100$ms, is farthest from the interpolations at $25/75/125/175$ms.

\begin{figure}
\centering
\includegraphics[width=.46\textwidth]{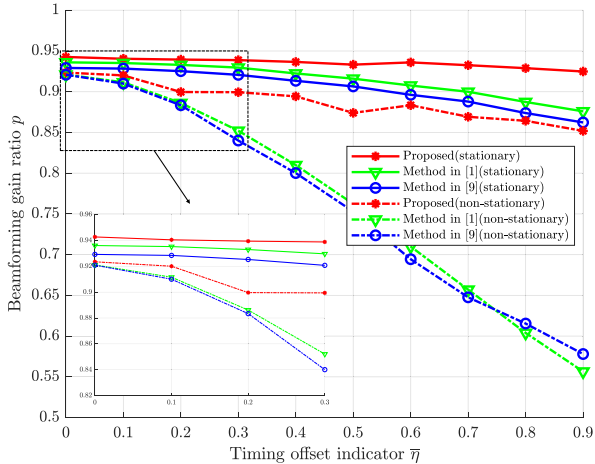}
\caption{Beamforming gain ratio comparison under different timing offset indicators.}
\end{figure}

The impact of Ricean K factor $R$ on beamforming gain ratio is investigated in Fig. 5. It is observed that both proposed scheme and deep learning based method in [9] show stronger robustness to $R$ than EKF based method in [1] in stationary scenarios, which demonstrates that deep learning models can extract robust features from complex environments with serious multipaths. Conversely, the performance of our proposed scheme enhances more rapidly than conventional methods as $R$ increases in non-stationary scenarios, since the baselines are difficult to process the highly nonlinear correlations.

\begin{figure}
\centering
\includegraphics[width=.49\textwidth]{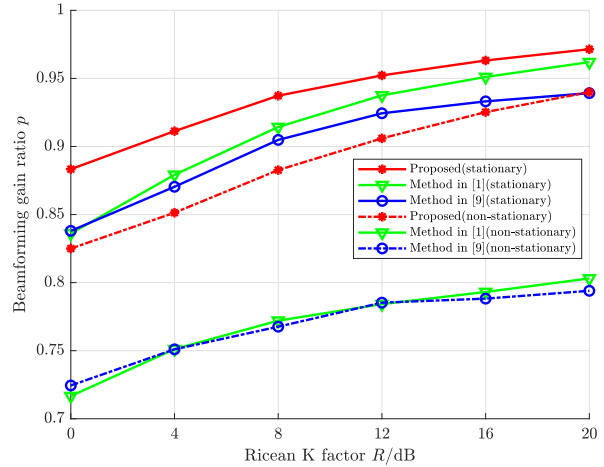}
\caption{Beamforming gain ratio comparison under different Ricean K factors.}
\end{figure}

\section{Conclusions}\label{sec:conclusion}


In this paper, a deep learning assisted beam prediction scheme, which utilizes the prior low-frequency CSI to predict the optimal mmWave beam, is proposed to reduce the overhead of mmWave beam training.
Specifically, timing offset indicator is proposed to indicate the instant of mmWave beam training relative to low-frequency CSI estimation.
Furthermore, grouped LSTMs is proposed to predict the optimal mmWave beams between periodical low-frequency CSI estimations, where the predicted result with the shortest time interval from mmWave beam training is used as the corresponding optimal beam.
Simulation results demonstrate that the proposed scheme can achieve higher beamforming gain compared with its conventional counterparts, especially under serious multipath interference and non-stationary scenarios.

\section*{Acknowledgment}

This work was supported in part by the National Key R\&D Program of China under Grant 2018YFB1801102, in part by National Natural Science Foundation of China (Grant No. 61871253), and in part by Postdoctoral Science Foundation of China under Grant 2020M670332. (\emph{Corresponding author: Zhaocheng Wang}).


\end{document}